\documentclass[pdflatex,sn-mathphys]{sn-jnl}

\theoremstyle{thmstyleone}%
\theoremstyle{thmstyletwo}%
\theoremstyle{thmstylethree}%

\usepackage{graphicx}
\usepackage{dcolumn}
\usepackage{bm}
\usepackage{hyperref}
\usepackage{xcolor}

\usepackage{lineno}
\begin{document}

\title[Catastrophic magnetic flux avalanches in NbTiN superconducting resonators]{Catastrophic magnetic flux avalanches in NbTiN superconducting resonators}


\author*[1]{\fnm{Lukas} \sur{Nulens}}\email{lukas.nulens@kuleuven.be joris.vandevondel@kuleuven.be}
\equalcont{These authors contributed equally to this work.}
\author[2]{\fnm{Nicolas} \sur{Lejeune}}
\equalcont{These authors contributed equally to this work.}
\author[1]{\fnm{Joost} \sur{Caeyers}}
\author[2]{\fnm{Stefan} \sur{Marinkovi\'c}}
\author[3]{\fnm{Ivo} \sur{Cools}}
\author[1]{\fnm{Heleen} \sur{Dausy}}
\author[1]{\fnm{Sergey} \sur{Basov}}
\author[1]{\fnm{Bart} \sur{Raes}}
\author[1]{\fnm{Margriet} \sur{J. Van Bael}}
\author[3]{\fnm{Attila} \sur{Geresdi}}
\author[2]{\fnm{Alejandro } \sur{V. Silhanek}}
\author[1]{\fnm{Joris} \sur{Van de Vondel}}

\affil*[1]{\orgdiv{Quantum Solid-State Physics, Department of Physics and Astronomy}, \orgname{KU Leuven}, \orgaddress{\street{Celestijnenlaan 200D}, \city{Leuven}, \postcode{B-3001}, \country{Belgium}}}
\affil[2]{\orgdiv{Experimental Physics of Nanostructured Materials, Q-MAT, CESAM}, \orgname{Université de Liège}, \orgaddress{\street{Allée du 6 Août 19}, \city{Sart Tilman}, \postcode{B-4000}, \country{Belgium}}}
\affil[3]{\orgdiv{Quantum Device Physics Labaratory, Department of Microtechnology and Nanoscience}, \orgname{Chalmers University of Technology}, \city{Goteborg}, \postcode{SE-412 96},  \country{Sweden}}


\abstract{The impact of the magnetic field penetration on the resonance frequency of NbTiN superconducting resonators is investigated by a combination of magneto-optical imaging and high-frequency measurements. At temperatures below approximately half of the superconducting critical temperature, the development of magnetic flux avalanches manifests itself as jumps in the resonance frequency of the coplanar resonators. A clear change in the rate of decreasing resonance frequency with magnetic field is observed when a magnetic perforation event regime sets the transition between a Meissner-like phase in the ground plane and flux injection into the central feedline. These regimes are directly visualized by magneto-optical imaging and the impact of avalanches in the ground plane and the resonator is discerned. At high working temperatures, a smooth flux penetration is shown to prevail. We propose some hints and strategies to mitigate the influence of avalanches on the response of the resonators and to improve the magnetic resilience of coplanar resonators. Our findings demonstrate that superconducting resonators represent a valuable tool to investigate the magnetic flux dynamics in superconducting materials. Moreover, the current blooming of niobium-based superconducting radio-frequency devices makes this report timely by unveiling the severe implications of magnetic flux dynamics.}




\maketitle

\section{Introduction}\label{sec1}

    Superconducting coplanar waveguide (CPW) resonators have become an essential component of quantum circuits due to their ability to readout different qubit systems. These CPW resonators combine a conventional fabrication method with superior quality factors needed to perform circuit quantum electrodynamics. During the last decade, careful design and material selection has resulted in quality factors reaching values up to $10^6-10^7$ \cite{paper_megrant2012planar,paper_vissers2010low}.   In order to obtain this high performance, the CPW resonators must be screened from external damping sources among which magnetic flux quanta play a particularly detrimental role. Although efficient magnetic screening can be achieved, this is not always a viable option since some qubit implementation schemes such as spin ensembles in solid-state systems \cite{paper_kubo2010strong,paper_schuster2010high,paper_amsuss2011cavity}, phase-slip qubits \cite{paper_astafiev2012coherent,paper_peltonen2018hybrid,paper_mooij2005phase,paper_mooij2006superconducting}, and trapped electrons \cite{paper_schuster2010proposal,paper_bushev2011trapped}, require inevitable exposure to a magnetic field. For these systems, it is crucial that the resonator characteristics remain unaffected under an external magnetic field. Unfortunately, previous studies have shown that increasing the magnetic field leads to a decrease in the quality factor as well as the resonance frequency of the resonator \cite{Song_resonator_field2009, paper_bothner2011improving,paper_bothner2012reducing,paper_bothner2012magnetic, paper_chiaro2016dielectric,paper_kroll2019magnetic}. Consequently, the identification of new strategies and designs for increasing the magnetic field resilience of superconducting CPW resonators remains a subject of major technological significance.
    \\
    \noindent
    The deleterious effect of magnetic field on the resonator characteristics can be attributed to the penetration of magnetic flux quanta or vortices \cite{Raes2012_singlevortexmotion}. Their interaction with the induced radio-frequency (RF) currents gives rise to the absorption of energy and consequently dissipation which translates into a decrease in the resonator performance. This indicates that not only the amount of vortices penetrating the CPW resonator is relevant but also their exact location.  In particular, the presence of vortices should be avoided in regions of maximal RF currents and/or magnetic field \cite{Nsanzineza_impact_singlevortex2014}. Two approaches can be adopted with the aim to lessen vortex-induced losses: (i) hinder their movement, and (ii) decrease their number in the critical regions. Several works have focused on the first approach by creating artificial pinning sites \cite{paper_bothner2011improving,paper_bothner2012reducing,paper_bothner2012magnetic,paper_song2009reducing,paper_kroll2019magnetic}. Alternatively, it has been shown that reducing the ground plane \cite{paper_bothner2017improving}, or dividing the ground plane into fractal structures \cite{paper_graaf2012magnetic,paper_de2014galvanically} permits to decrease the demagnetization effects and decrease the field penetration.

 
In order to seize and quantify the impact of the aforementioned mitigation strategies, the complete sample surface should be taken into account. To that end, a macroscopic technique capable of visualizing the magnetic flux penetration and structural weak points is essential. Magneto-optical imaging (MOI) has proven to be an excellent tool for examining macroscopic structures such as superconducting CPW resonators with $\mu$m spatial resolution \cite{paper_lange2017high,paper_ghigo2007evidence}. In order to make morphometric criteria-based decisions for field-resilient CPW resonators, the macroscopic image of a full device must be complemented and compared to high-frequency transmission measurements.

   In this work, we investigate the effect of the magnetic field penetration on a superconducting CPW resonator. To that end, we designed a sample containing two $\lambda/4$ resonators able to be characterized by magneto-optical imaging and high-frequency measurements at several temperatures. Our results show that both observed resonances follow the expected temperature behavior determined by the kinetic inductance of the material. We analyzed the field dependence of the high-frequency measurements and observed several different regimes in the behavior of the resonance frequency. These regimes were directly correlated to the MOI observations, with a particular focus on the noisy behavior in the resonance frequency that can be attributed to the development of thermomagnetic instabilities precursors of abrupt flux avalanches. Based on our findings, we suggest several ways to optimize the field resilience of future CPW resonators.

\section{Results and Discussion}\label{sec2}

\subsection{Sample design}\label{sec3}
\begin{figure*}[htpb]
\centering
\includegraphics[width=\linewidth]{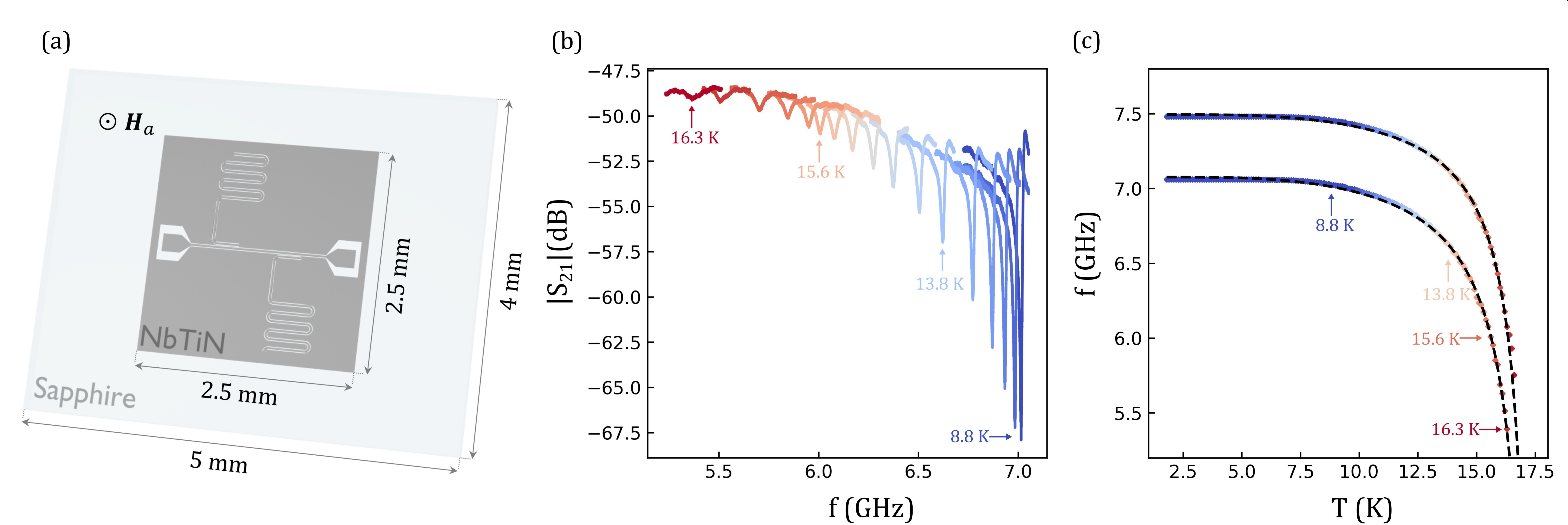}
\caption{(a) A schematic representation of the investigated sample. In the superconducting NbTiN region indicated in dark grey, two $\lambda/4$ resonators are capacitively coupled to a central feedline. (b) The $S_{21}$ transmission parameter of the longest resonator plotted for different fixed temperatures between 8.8 K and 16.3 K. The color change from blue to red indicates the increase in temperature. (c) For both resonators indicated in panel a the resonance frequency as a function of the temperature is indicated from blue to red with increasing temperatures. The dashed black line corresponds to a fit of both resonance frequencies as described in the text. The data points corresponding to the transmission measurement shown in panel b are indicated with the corresponding colored arrow.} 
\label{fig_1}
\end{figure*}

 The superconducting resonators investigated in this work consist of a 100 nm thick NbTiN layer sputtered on a sapphire substrate, as schematically represented in Figure \ref{fig_1}(a). A critical temperature of 16.5 K was obtained using SQUID magnetometry on a reference layer of 100 nm. Several identical samples were fabricated in order to allow comparison between the MOI and RF measurements. In order to fit the complete resonator structure in the field of view of the MOI setup a substrate size of $(5 \times 4) \text{ mm}^2$ with a superconducting region of interest of $(2.5 \times 2.5) \text{ mm}^2$ was chosen. In this superconducting region two overcoupled, hanger-type $\lambda/4$ resonators with length 4089 $\mu m$ and 3953 $\mu m$ were capacitively coupled to a central feedline. The central conductor has a width $w$ = 20 $\mu$m with a separation from the groundplane of $s$ = 10 $\mu$m resulting in a 50 $\Omega$ impedance. 
 
\subsection{Resonator characteristics}\label{sec3}

 The resonance frequencies, $f_r$, at 7.06 GHz and 7.44 GHz corresponding to each resonator manifest themselves as an absorption dip in the $S_{21}$ transmission spectrum at 1.8 K. The resonance dips showed a pronounced skewness likely due to an impedance mismatch caused by the lengthy wire bonds ($\sim$ mm) required to fit the high-frequency setup \cite{paper_doyle2008lumped,paper_grabovskij2008situ,paper_khalil2012analysis}. In Figure \ref{fig_1}(b) the magnitude of the transmission is shown for different temperature values. As the temperature increases from lower (blue) to higher (red) values, the resonance frequency shifts to lower values and the full width at half maximum increases, indicating an increase in losses. Figure \ref{fig_1}(c) displays the temperature dependence of both resonance features. The resonance frequencies 7.06 GHz and 7.44 GHz with corresponding internal quality factors of $\sim 2000 $ and $\sim 800$ were obtained following the fitting routine described in Ref. \cite{paper_probst2015efficient}. \\

In general, the resonance frequency of a $\lambda/4$ resonator is determined by the expression $f_r = 1/(4l\sqrt{L'C'})$, where $l$ corresponds to the resonator length, and $L'$ and $C'$ are the inductance and capacitance per unit length, respectively \cite{Book_pozar2011microwave}. The resonance frequencies were originally designed based solely on a geometric inductance contribution of $L'_g = 4.19\cdot10^{-7}$ H/m. However, a kinetic inductance contribution should also be taken into consideration. Specifically, this kinetic inductance can be expressed as a function of the temperature-dependent superconducting energy gap, $\Delta(T)$ \cite{Book_tinkham2004introduction,paper_annunziata2010tunable}:

         \begin{equation}
         L'_k(T) = \frac{l}{w}\frac{R_{\square}h}{2\pi^2\Delta(T)}\frac{1}{tanh\left(\frac{\Delta(T)}{2k_BT}\right)} ,
         \label{eq_Lk}
         \end{equation}

    with $l$ and $w$ corresponding to the length and width of the resonator, $R_{\square}$ the normal state sheet resistance, $h$ and $k_B$ are Planck's and Boltzmann's constants, and $T$ is the temperature \cite{Book_tinkham2004introduction,paper_annunziata2010tunable}. By inserting an interpolation formula for the temperature-dependent superconducting energy gap $\Delta(T) \approx \Delta(0)\tanh(1.74\sqrt{T_c/T-1})$ following the BCS-theory in Equation \ref{eq_Lk}, the temperature dependent resonance frequency can be fitted with fitting parameters $L'_k(T=0), T_c$ and $L'_g$. The fit for both resonators is indicated in dashed black lines in Figure \ref{fig_1}(c), with parameters $L'_g = 3.82\cdot 10^{-7}$ H/m, $L'_k = 3.73 \cdot 10^{-8}$ H/m, $T_c = 17.27$ K and $L'_g = 4.16\cdot 10^{-7}$ H/m, $L'_k = 5.46 \cdot 10^{-8}$ H/m, $T_c = 17.21$ K for the shortest and longest resonator respectively. The excellent fit demonstrates that the resonance frequency serves as a very sensitive probe for any variation of the superconducting energy gap. \\

\subsection{Magnetic field response of the resonators}\label{sec4}

\begin{figure*}[htpb]
		\centering
            \includegraphics[width=\linewidth]{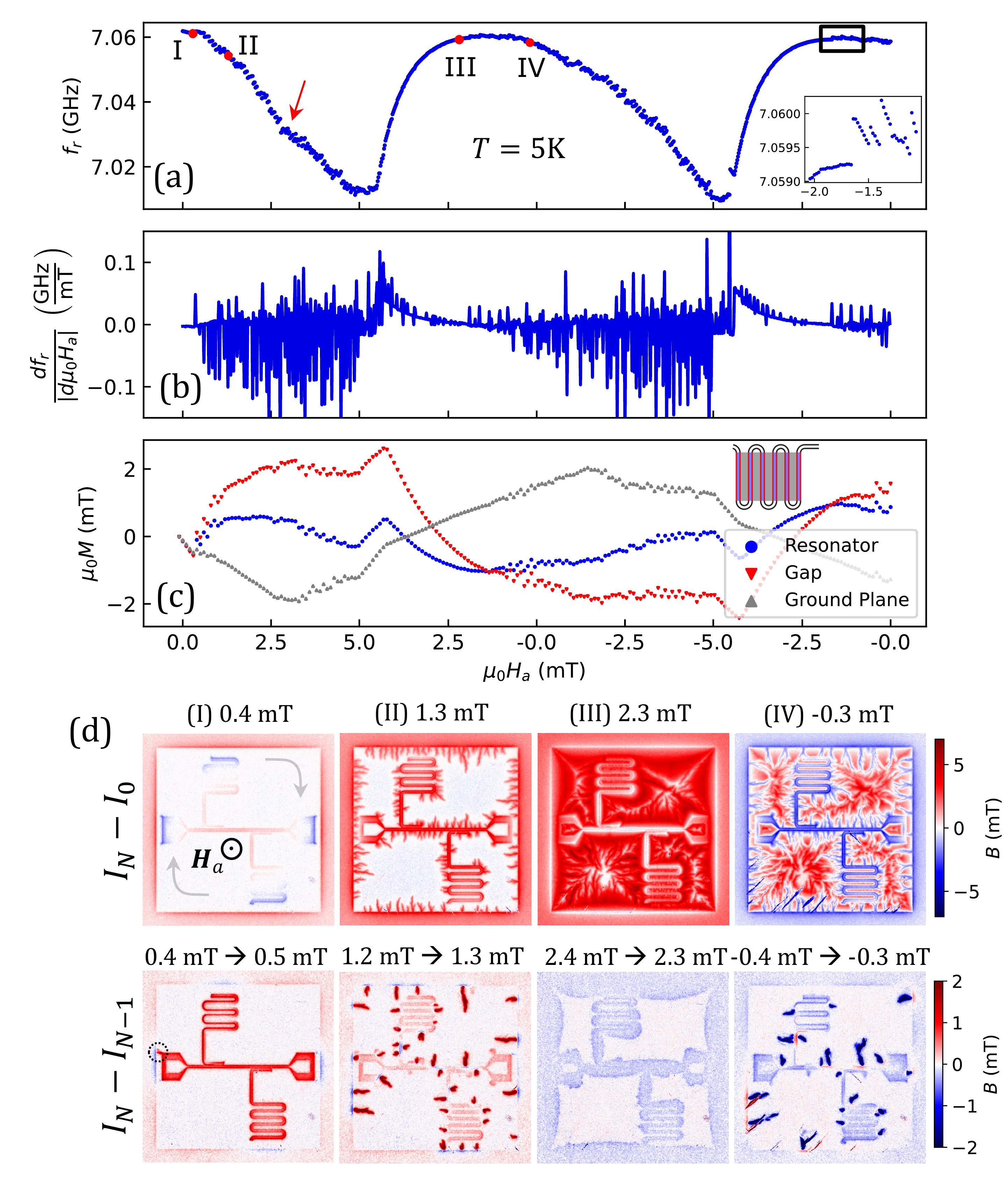}
  
		\caption{(a) Resonance frequency as a function of the applied magnetic field $\mu_0 H_a$ obtained at $T=5$ K. The magnetic field has been swept to complete a full loop $0 \rightarrow 5 \rightarrow -5 \rightarrow 0$ mT. The inset shows a zoom-in of the framed region. (b) The derivative $df_r/d\mu_0 H_a$ extracted from panel a makes the noisy and smooth regime apparent. (c) The local magnetization as a function of the applied magnetic field inside the region schematically represented in the inset. The blue, red, and grey data points correspond with the magnetization averaged over the central conductor of the meander, the gap separating the central conductor and the ground plane, and the ground plane (pitch) between the meander structure respectively. (d) The upper row of the panel shows magneto-optical images corresponding to the magnetic field indicated in panel a. The Meissner currents are indicated with grey arrows in the image corresponding to point I. The lower row of the panel shows differential images revealing the change of magnetic flux between two consecutive field steps. The nucleation point where the magnetic perforation takes place is indicated with a dotted circle.}
		\label{fig2}
	\end{figure*}

Let us investigate the response of the resonators as a function of the applied out-of-plane magnetic field ($\mu_0H_a$). Figure \ref{fig2}(a) shows the resonance frequency of the longest resonator after a zero-field cooling process and subsequent sweeping of the magnetic field at the fixed temperature of 5 K. For the sake of clarity, we have unfolded the magnetic field cycle in such a way that the abscissa can be associated to the measurement chronology. Note that as magnetic field increases, the resonant frequency shifts down with an average rate smaller than the subsequent increase rate of $f_r$ when decreasing field. In other words, $f_r(\mu_0 H_a)$ exhibits an irreversible response when cycling the applied magnetic field. This hysteric behavior of $f_r$ is a universal feature which has been shown for Nb \cite{paper_bothner2012magnetic,paper_bonura2006magnetic}, NbN \cite{paper_yu2021magnetic}, YBCO \cite{paper_lahl2003nonlinear}, MgB$_2$ \cite{paper_ghigo2010mechanisms}, and Al \cite{paper_borisov2020superconducting} CPW resonators and can be linked to the inhomogeneous distribution of current, $I_{rf}$, shaking an inhomogeneous magnetic flux distribution. As shown in the supplementary figure 1 the internal quality factor has a very similar magnetic field dependence. This demonstrates that the same interplay defines dissipation in these resonators. 

\noindent

A close inspection of Figure \ref{fig2}(a) allows us to identify several different features which should be associated with the peculiarities of the magnetic field penetration in the complete structure. As we will demonstrate below, in order to identify the non-local magnetic origin of these features it is essential to map the magnetic field distribution in the entire device (shown in Figure \ref{fig2}(d)).   

Figure \ref{fig2}(a) shows that the resonance frequency maximizes at $\mu_0 H_{a}=0$ mT, where little magnetic flux (mainly due to frozen earth magnetic field) remains trapped in the resonators. In the field range $0 \text{ mT} < \mu_0 H_a < 0.4$ mT, screening Meissner currents circulate in the ground plane (point (I) in Figure \ref{fig2}(a,d)). Note that the ground plane and the resonators are capacitively coupled to the central excitation line. In other words, the ground plane and resonators are physically disconnected from the excitation line and surround it. Therefore, the structure can be regarded as topologically equivalent to a ring surrounding a disk. Due to demagnetization effects, the Meissner currents circulating clockwise in the ground plane (see grey arrows in image (I) of Figure \ref{fig2}(d)), produce a magnetic field of opposite polarity than the applied field all along the inner border of the ground plane and a small positive field inside the central hole (where the feed line sits). As long as $0 \text{ mT} < \mu_0 H_a < 0.4$ mT the resonators are protected by the screening currents in the ground plane and $f_r$ gently decreases as magnetic field increases.
\\
\noindent
At $\mu_0 H_a = 0.5$ mT a peculiar event, known as magnetic perforation takes place. This event consists of a sudden breakdown of the screening currents in the ground plane with a simultaneous injection of magnetic flux inside the central hole \cite{paper_olsen2007avalanches,paper_shvartzberg2019quasiperiodic,paper_jiang2020selective}. This abrupt change of the magnetic field at the rim of the resonators marks the start of a steeper magnetic field dependence of $f_r$. The injection of the magnetic field becomes more apparent in the differential image (i.e. subtraction of two images taken a two consecutive magnetic fields) shown in Figure \ref{fig2}(d) corresponding to point (I) in which the nucleation point of the magnetic perforation has been indicated with a dotted circle. \\
As the magnetic field is progressively increased beyond the magnetic perforation point ($0.5 \text{ mT}< \mu_0 H_a < 5$ mT), $f_r(\mu_0H_a)$ exhibits a noisy behavior with a global linear decrease as the magnetic field increases. This noisy regime is better seen in the derivative $df_r/d\mu_0 H_a$ shown in Figure \ref{fig2}(b). The origin of this jumpy behavior can be traced back to the development of magnetic flux avalanches caused by thermomagnetic instabilities \cite{RevModPhys_criticalstate}. A snapshot of the magnetic landscape after triggering some finger-shaped flux avalanches at the point labeled (II) is shown in Figure \ref{fig2}(d). These avalanches are neither fully reproducible \cite{paper_johansen2007reproducible,paper_qureishy2017dendritic}
nor totally random and could equally originate at border spots with high degree of perfection or defects \cite{paper_brisbois2016magnetic,paper_jiang2020selective}. 


When decreasing the magnetic field from $\mu_0 H_a = 5$ mT, the local magnetic field inverts its polarity at the border of the sample even before the applied field changes polarity \cite{Brandt1993_edgeH, Zeldov1994_edgeH,McDonald1996_edgeH}. In this case, antivortices nucleate at the border of the sample and annihilate with the trapped vortices. At the beginning, this process takes place smoothly and leads to an effective removal of magnetic flux from the resonators. Since no flux avalanches occur in this regime, the $f_r(\mu_0H_a)$ curve does not exhibit any jumpy behavior (see point (III) in Figure \ref{fig2}(a,d)). The fact that the flux is efficiently removed from the border of the resonators due to the pronounced decrease of the total magnetic field at the sample's borders gives rise to a lower density of vortices when decreasing field than for increasing field for the same applied field \cite{Brandt1993_edgeH, Zeldov1994_edgeH,McDonald1996_edgeH}. This irreversible behavior, in turn, leads to a steeper increase of $f_r(\mu_0H_a)$ as $\mu_0H_a$ decreases as compared to the up sweep branch. Note that in this regime flux avalanches are no longer triggered although those created previously leave a frozen imprint on the magnetic landscape and may act as channels for the penetration of antivortices \cite{paper_motta2011visualizing}. The differential imaging shown in Figure \ref{fig2}(d)-(III) clearly demonstrates that no avalanches develop in this regime. 
\\
\noindent
The smooth regime finishes at a certain field $\mu_0 H^* \approx 2.5$ mT. Further decreasing the applied field beyond this point, triggers avalanches of flux with opposite polarity to the applied field. This is naturally accompanied by a jumpy $f_r(\mu_0H_a)$ response all the way down to $\mu_0 H_a = -5$ mT. Snapshots of the actual magnetic landscape and the differential imaging for point (IV) within this regime are shown in Figure \ref{fig2}(d). A zoom of the transition from smooth to jumpy evolution of $f_r(\mu_0H_a)$ is shown in the inset of Figure \ref{fig2}(a), corresponding to the framed area. Note here that abrupt jumps which tend to increase $f_r$ are followed by smooth linear regions where $f_r$ decreases. This behavior can be explained as follows: after the event of a flux avalanche, the magnetic field and current at the border of the resonator are relaxed, and therefore $f_r$ increases. As $\mu_0H_a$ increases further, the system builds up magnetic pressure (local field and current increases at the border of the resonator), and therefore $f_r$ slowly decreases.
\\
\noindent
It is interesting to note the presence of a kink in the vicinity of the middle zone of the noisy regime in panel (a), indicated by a red arrow. In order to identify the origin of this feature, we have measured the local magnetic field $B$ in different regions of the device and computed the local magnetization $\mu_0M=B-\mu_0H_a$. The resulting magnetization $\mu_0M$ as a function of the applied field is shown in Figure \ref{fig2}(c) for the resonator (blue dots), the gap separating the resonator from the ground-plane (red downwards triangles), and the ground plane in between the meander shape resonator (grey upwards triangles). These regions are indicated with the same color code in the inset of Figure \ref{fig2}(c). This figure shows that the kink in the $f_r(H)$ curve arises at the same moment when the local magnetization of the ground plane starts to increase as a consequence of the development of magnetic flux avalanches in the ground plane. These events relieve the magnetic pressure everywhere in the resonator and tend to slow down the decrease of the resonance frequency. A selection of MOI images of the resonator meander showing the point avalanches is available in Supplementary Figure 2.
\\
\noindent

The magnetic and magnetization profiles over the magnetic sweep of the region indicated in the inset of Figure \ref{fig2}(c) can be found in supplementary video 1. Similarly, the evolution of the $B(\mu_0 H_a)$ curve as a function of the measurement location along the same region can be found in supplementary video 2.

\subsection{Temperature response of the resonators}\label{sec5}
     	 
\begin{figure*}[!htbp]
		\centering
		\includegraphics[width=\linewidth]{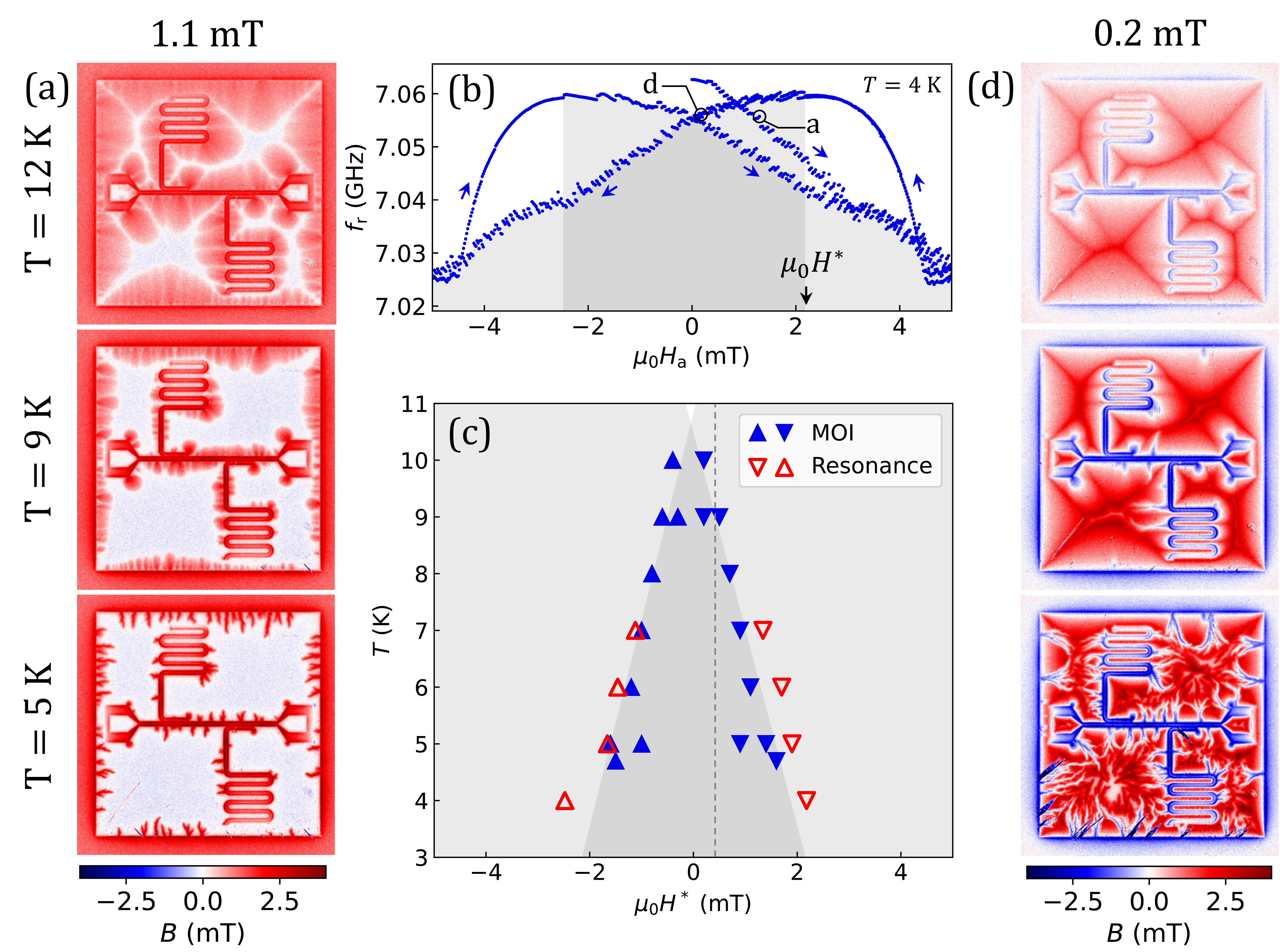}
		\caption{Temperature dependence of the magnetic flux avalanche activity. (a) Magnetic landscape after zero-field cooling at three different temperatures (5, 9 and 12 K) for the same applied field $\mu_0 H_a=1.1$ mT. (b) Hysteresis loop of the resonance frequency $f_r(\mu_0H_a)$ obtained at 4 K. The region where (anti)flux avalanches are triggered when decreasing (increasing) from a maximum (minimum) applied field, are highlighted with a grey background. (c) Threshold field $\mu_0H^*$ as a function of temperature as determined by the onset of the noisy regime in $f_r(\mu_0H_a)$ measurements (red triangles) and by magneto-optical imaging (blue triangles). The dotted line indicates the field of 0.2 mT at which the images in panel (d) have been selected. The experiment has been conducted twice at 5 K and 9 K resulting in additional blue ($\bigtriangledown$,$\bigtriangleup$) datapoints. (d) Magnetic landscape after zero-field cooling at three different temperatures (5, 9 and 12 K) for the same applied field $\mu_0 H_a=0.2$ mT. }
	\label{fig3}
\end{figure*}

As shown in the previous section, flux avalanches have a critical impact on the overall performance of the CPW resonators. In Figure \ref{fig3}  we identify the transition from the flux avalanche regime to smooth magnetic penetration using both the magnetic field dependent resonance frequency and the MOI  images. Panel (a) shows the magnetic flux penetration after zero-field cooling for three different temperatures (5, 9, and 12 K) at the same applied field $\mu_0 H_a=1.1$ mT. Finger-like avalanches clearly visible at 5 K are no longer present at 9 and 12 K. Indeed, it is well known that the occurrence of thermomagnetic instabilities is restricted to low temperatures (roughly for $T<T_c/2$).  At $T=9$ K, it is observed that border defects act as magnetic faucets from where magnetic flux is injected into the ground plane, a phenomenon which has been described in detail in Ref.\cite{paper_brisbois2016magnetic}.
At $T=12$ K, the sample has been largely penetrated by the magnetic field and as the magnetic field increases, non-penetrated white regions shrink and converge to form discontinuity lines where the supercurrents change direction \cite{paper_gurevich2000nonlinear}.

As we have pointed out above, the magnetic flux avalanche activity becomes more apparent when decreasing the applied field below certain threshold value $\mu_0H^*$ after reaching full penetration. This effect is a direct consequence of the high energy release produced by the vortex-antivortex annihilation which facilitates the development of avalanches with opposite polarity than the applied field. The threshold field $\mu_0H^*$ at which anti-avalanches are triggered can be identified either as the onset of jumpy $f_r(\mu_0H_a)$ response (shadowed region in panel (b)) or can be directly obtained from the magneto-optical images. Figure \ref{fig3}(c) shows the threshold field $\mu_0H^*$ as a function of temperature obtained by these two independent methods for the decreasing ($\bigtriangledown$) and increasing ($\bigtriangleup$) branches of the hysteresis loops. To ensure consistency, we repeated the experiment during a second cooldown at 5 K and 9 K. As a result, Figure \ref{fig3}(c) displays multiple data points for these temperatures. As $T$ increases $\mu_0H^*$ progressively diminishes showing no magnetic flux avalanche activity above 10 K. Panel (d) shows MOI images acquired in the decreasing branch of the hysteresis loop at $\mu_0 H_a=0.2$ mT which corresponds to the dotted line shown in panel (c) and illustrating the presence of antiflux avalanches at 5 and 9 K and their absence at 12 K. 

A complete animation of the evolution of the magnetic landscape during the magnetic loop can be found in the supplementary video 3.

\section{Conclusion}\label{sec13}

We explored NbTiN CPW resonator structures using both magneto-optical imaging and RF transmission experiments. A clear and unambiguous correspondence is demonstrated between both independent methods.  This allows us to directly link the complex magnetic field penetration to the changes in the resonance frequency in an applied magnetic field. Major attention is drawn to magnetic flux avalanches, which drastically impact the resonance frequency of the device when triggered. Based on the measurements and analysis presented above, we can devise a roadmap for design optimization of the superconducting resonators with the goal of improving their resilience and performance under perpendicular magnetic fields:

\begin{itemize}
  \item {\it Ground plane}. In order to increase the threshold field needed for magnetic perforation, it is essential to decrease the local field at the outer rim of the ground plane. In the Meissner state this local field can be approximated \cite{paper_zeldov1994magnetization} as  $\approx \sqrt{W/t} \mu_0 H_a$, where $W$ is the width of the ground plane and $t$ is the thickness of the superconducting film. Therefore, the desired effect can be achieved by reducing the lateral dimension $W$ of the ground plane. 
  \item {\it Resonators}. If flux avalanche-induced noise needs to be suppressed, a possible solution consists in reducing the pitch of the meanders. Indeed, it has been shown in Ref. \cite{paper_denisov2006onset} that thermomagnetic instabilities are progressively suppressed as the width of the superconductor decreases and are fully absent below certain threshold value. This is similarly exemplified in the field resilience of nanowire NbTiN resonators \cite{paper_samkharadze2016high}. Alternatively, by introducing artificial roughness along the borders of the superconducting device it should be possible to shrink the $H-T$ region where avalanches occur \cite{paper_cerbu2013vortex,paper_brisbois2016magnetic}.  
  \item {\it Artificial defects}. It has been demonstrated that the introduction of periodic pinning such as through holes can be beneficial for reducing the losses caused by the flux shaking. However, this may only hold at high enough temperatures where flux avalanches do not take place. Indeed, it is known that this kind of pinning may actually favor the triggering of magnetic flux avalanches and deteriorate the performance of the device at low temperatures \cite{paper_menghini2005dendritic,paper_motta2014controllable}. Another interesting approach consists in adding pinning centers in the resonator there where the RF local field maximizes, as proposed in Ref. \cite{paper_ghigo2004microwave}.
  \item {\it Ratchet potential}. It may be worth exploring the possibility to keep the magnetic flux at bay by introducing asymmetric pinning landscapes forming flux lenses \cite{paper_lee1999reducing}.
  Although the efficiency of this sort of ratchet potentials have been demonstrated at relatively low frequencies, their performance at GHz seem to be less promising \cite{paper_dobrovolskiy2020upper}.   
\end{itemize}

It is interesting to mention that some high-$T_c$ superconductors such as YBa$_2$Cu$_3$O$_7$ do not exhibit much flux avalanche activity, at least at field rates normally used in the laboratory. However, 
the possibility exists that the rapidly oscillating RF field and the associated huge rate of change stimulate the otherwise absent flux avalanches in these materials. Therefore, it would be interesting to extend the present work to perform MOI and frequency measurements simultaneously. This endeavour risks to be highly non-trivial since the proximity of the Al mirror and the magnetic indicator needed for MOI measurements may substantially distort the performance of the resonators. If nevertheless one succeeds in this task, it will be possible to discern if the RF excitation itself promotes or stimulates magnetic flux avalanches \cite{paper_awad2011flux}.   
 
\section{Methods}\label{sec11}
\subsection{Sample fabrication}\label{secM3}
The resonators were fabricated on a double side polished (430 $\pm$ 25) $\mu$m thick C-plane Al$_2$O$_3$ wafer. First, the wafer was cleaned using a standard 5:1:1 RCA process of deionized water, ammonia water, and H$_2$O$_2$ respectively and heated to 80 $ ^{\circ}$C for 10 minutes \cite{Kirby2008}. Subsequently, one side of the wafer was covered by 100 nm NbTiN by sputtering on a target of Nb$_{81.9}$Ti$_{18.1}$N \cite{Burton2015} in a near-UHV sputter system, under an Ar:N$_2$ partial gas pressure ratio of 60:37. Prior to, and during deposition the wafer was annealed at 660 $ ^{\circ}$C for 30 minutes. Subsequently, the coplanar waveguide resonators are patterned on the sample side using a PMMA A4 mask, pre-baked at 180 $ ^{\circ}$C. An e-spacer 300Z conduction layer is applied to mitigate charging issues that may arise. After development in IPA:H$_2$O 4:1, an Ar/Cl (4:50 sccm) ICP plasma etches the metal \cite{Mahashabde2020,Niepce2020}. Finally, the remainder of the resist was removed by a heated aceton.
\subsection{RF measurements}\label{secM1}
    High-frequency measurements were performed using a Keysight P5003B Streamline Vector Network Analyzer in a Janis $^3$He cryostat. The input line was attenuated with 60dB, which was distributed across different temperature stages within the cryostat. The sample was pasted onto an oxygen-free copper holder and connected via wire bonding to an RF holder with a sample space of (7 × 4) mm$^2$. A 4-8 GHz circulator/isolator and 4-8 GHz high-electron-mobility transistor (LNF-LNC4\textunderscore8C) were connected at the 4K stage after the output line of the sample, amplifying the signal back to the second port of the VNA. The magnetic field was applied perpendicular to the sample using a superconducting magnet, as illustrated in Figure \ref{fig_1}a. To compensate for the remanent field, the sample was field cooled ($-0.2$ mT) before every measurement cycle. A measurement cycle consists of sweeping the magnetic field as follows : $0 \rightarrow 5 \rightarrow -5 \rightarrow 5 \rightarrow 0$ mT completing a full loop. Initial field cycles were executed with a driving power of $- 40$ dBm and compared to a higher driving power of $0$ dBm. As there was no significant difference all measurements were executed at  $0$ dBm.
    \\
   \subsection{MOI technique}\label{secM2}
    Direct visualization of the magnetic flux landscape was obtained by magneto-optical imaging. This technique is based on the Faraday rotation of linearly polarized light in a $3~\mu$m-thick Bi-doped yttrium iron garnet (indicator) with in-plane magnetic domains, placed on top of the investigated sample \cite{paper_koblischka1995magneto}. 
    Since the rotation of polarization is proportional to the local magnetic field $B_z$ at the indicator, the use of an analyzer oriented perpendicularly to the initial direction of polarization results in images where the intensity is proportional to $B_z$. The images are acquired with a CCD camera and have a pixel size of $1.468 \times 1.468~\mu$m$^2$. Post-image processing was done to remove the inhomogeneous illumination and field-independent background, using the ImageJ software. More information about the MO imaging setup can be found in Ref. \cite{paper_shaw2018quantitative}.
    Low temperature MO measurements are performed in a closed-cycle cryostat and the external magnetic field was applied through a copper coil.  MO imaging allows to record spatial maps of the magnetic flux and thus obtain direct information on the location and size of each event. 
\\
\subsection{SQUID magnetometry}
Magnetic measurements were carried out in a superconducting quantum interference device magnetometer (Quantum Design SQUID VSM MPMS3). The DC magnetization versus temperature was measured in a parallel applied field upon heating the sample from 1.8 to 300 K under a remanent magnetic field in the order of 0.01 mT. The critical temperature $T_c$ was defined as the highest temperature with a measurable diamagnetic magnetization contribution on top of the linear background. A demagnetization procedure was conducted at 300 K prior to the measurement by oscillating the field from -7 Tesla to near zero to minimize the remanent magnetic field from the superconducting magnet \cite{paper_blunt1991investigation}.

\section{Data availability}\label{sec11}
The data that support the plots of this paper and other findings within this study are available from the corresponding author upon reasonable request.

\section{Code availability}\label{sec11}
The codes used for this study are available from the corresponding author upon reasonable request.

\backmatter

\bmhead{Acknowledgments}
The authors acknowledge support from the EU COST action SUPERQUMAP CA21144, the Fonds de la Recherche Scientifique - FNRS under the grants Weave-PDR T.0208.23 and CDR J.0176.22, the Research Foundation Flanders (FWO, Belgium), Grants No. G0A0619N and 11K6523N, and the KU Leuven C1 program C14/18/074. N. L. acknowledges support from FRS-FNRS (Research Fellowships FRIA) and S. M. acknowledges support from FRS-FNRS (Research Fellowships ASP). I. C. and A. G. acknowledge support by the European Union's H2020 research and innovation program, grant no. 804988 (SiMS) and 828948 (AndQC). The authors thank Christian Haffner for his valuable support regarding the experimental development of the high-frequency setup.

\bmhead{Author contributions}
{L.N. and B.R. designed the devices, I.C. and A.G. fabricated the devices, N.L. and S.M. executed MOI measurements, L.N., J.C. and H.D. executed RF measurements, S.B. executed SQUID magnetometry. The manuscript was written by L.N., N.L., A.V.S. and J.V.V. with the help from all other authors. All authors discussed the results and reviewed the manuscript. J.V.V., M.J.V.B., B.R. and A.S. initiated and supervised the research.}

\bmhead{Competing interests}
The authors declare no competing interests.





\bibliography{Ref.bib}


\begin{thebibliography}{66}
\ifx \bisbn   \undefined \def \bisbn  #1{ISBN #1}\fi
\ifx \binits  \undefined \def \binits#1{#1}\fi
\ifx \bauthor  \undefined \def \bauthor#1{#1}\fi
\ifx \batitle  \undefined \def \batitle#1{#1}\fi
\ifx \bjtitle  \undefined \def \bjtitle#1{#1}\fi
\ifx \bvolume  \undefined \def \bvolume#1{\textbf{#1}}\fi
\ifx \byear  \undefined \def \byear#1{#1}\fi
\ifx \bissue  \undefined \def \bissue#1{#1}\fi
\ifx \bfpage  \undefined \def \bfpage#1{#1}\fi
\ifx \blpage  \undefined \def \blpage #1{#1}\fi
\ifx \burl  \undefined \def \burl#1{\textsf{#1}}\fi
\ifx \doiurl  \undefined \def \doiurl#1{\url{https://doi.org/#1}}\fi
\ifx \betal  \undefined \def \betal{\textit{et al.}}\fi
\ifx \binstitute  \undefined \def \binstitute#1{#1}\fi
\ifx \binstitutionaled  \undefined \def \binstitutionaled#1{#1}\fi
\ifx \bctitle  \undefined \def \bctitle#1{#1}\fi
\ifx \beditor  \undefined \def \beditor#1{#1}\fi
\ifx \bpublisher  \undefined \def \bpublisher#1{#1}\fi
\ifx \bbtitle  \undefined \def \bbtitle#1{#1}\fi
\ifx \bedition  \undefined \def \bedition#1{#1}\fi
\ifx \bseriesno  \undefined \def \bseriesno#1{#1}\fi
\ifx \blocation  \undefined \def \blocation#1{#1}\fi
\ifx \bsertitle  \undefined \def \bsertitle#1{#1}\fi
\ifx \bsnm \undefined \def \bsnm#1{#1}\fi
\ifx \bsuffix \undefined \def \bsuffix#1{#1}\fi
\ifx \bparticle \undefined \def \bparticle#1{#1}\fi
\ifx \barticle \undefined \def \barticle#1{#1}\fi
\bibcommenthead
\ifx \bconfdate \undefined \def \bconfdate #1{#1}\fi
\ifx \botherref \undefined \def \botherref #1{#1}\fi
\ifx \url \undefined \def \url#1{\textsf{#1}}\fi
\ifx \bchapter \undefined \def \bchapter#1{#1}\fi
\ifx \bbook \undefined \def \bbook#1{#1}\fi
\ifx \bcomment \undefined \def \bcomment#1{#1}\fi
\ifx \oauthor \undefined \def \oauthor#1{#1}\fi
\ifx \citeauthoryear \undefined \def \citeauthoryear#1{#1}\fi
\ifx \endbibitem  \undefined \def \endbibitem {}\fi
\ifx \bconflocation  \undefined \def \bconflocation#1{#1}\fi
\ifx \arxivurl  \undefined \def \arxivurl#1{\textsf{#1}}\fi
\csname PreBibitemsHook\endcsname

\bibitem{paper_megrant2012planar}
\begin{barticle}
\bauthor{\bsnm{Megrant}, \binits{A.}},
\bauthor{\bsnm{Neill}, \binits{C.}},
\bauthor{\bsnm{Barends}, \binits{R.}},
\bauthor{\bsnm{Chiaro}, \binits{B.}},
\bauthor{\bsnm{Chen}, \binits{Y.}},
\bauthor{\bsnm{Feigl}, \binits{L.}},
\bauthor{\bsnm{Kelly}, \binits{J.}},
\bauthor{\bsnm{Lucero}, \binits{E.}},
\bauthor{\bsnm{Mariantoni}, \binits{M.}},
\bauthor{\bsnm{O’Malley}, \binits{P.J.}}, \betal:
\batitle{Planar superconducting resonators with internal quality factors above
  one million}.
\bjtitle{Applied Physics Letters}
\bvolume{100}(\bissue{11}),
\bfpage{113510}
(\byear{2012}).
\doiurl{10.1063/1.3693409}
\end{barticle}
\endbibitem

\bibitem{paper_vissers2010low}
\begin{barticle}
\bauthor{\bsnm{Vissers}, \binits{M.R.}},
\bauthor{\bsnm{Gao}, \binits{J.}},
\bauthor{\bsnm{Wisbey}, \binits{D.S.}},
\bauthor{\bsnm{Hite}, \binits{D.A.}},
\bauthor{\bsnm{Tsuei}, \binits{C.C.}},
\bauthor{\bsnm{Corcoles}, \binits{A.D.}},
\bauthor{\bsnm{Steffen}, \binits{M.}},
\bauthor{\bsnm{Pappas}, \binits{D.P.}}:
\batitle{Low loss superconducting titanium nitride coplanar waveguide
  resonators}.
\bjtitle{Applied Physics Letters}
\bvolume{97}(\bissue{23}),
\bfpage{232509}
(\byear{2010}).
\doiurl{10.1063/1.3517252}
\end{barticle}
\endbibitem

\bibitem{paper_kubo2010strong}
\begin{barticle}
\bauthor{\bsnm{Kubo}, \binits{Y.}},
\bauthor{\bsnm{Ong}, \binits{F.}},
\bauthor{\bsnm{Bertet}, \binits{P.}},
\bauthor{\bsnm{Vion}, \binits{D.}},
\bauthor{\bsnm{Jacques}, \binits{V.}},
\bauthor{\bsnm{Zheng}, \binits{D.}},
\bauthor{\bsnm{Dr{\'e}au}, \binits{A.}},
\bauthor{\bsnm{Roch}, \binits{J.-F.}},
\bauthor{\bsnm{Auff{\`e}ves}, \binits{A.}},
\bauthor{\bsnm{Jelezko}, \binits{F.}}, \betal:
\batitle{Strong coupling of a spin ensemble to a superconducting resonator}.
\bjtitle{Physical Review Letters}
\bvolume{105}(\bissue{14}),
\bfpage{140502}
(\byear{2010}).
\doiurl{10.1103/PhysRevLett.105.140502}
\end{barticle}
\endbibitem

\bibitem{paper_schuster2010high}
\begin{barticle}
\bauthor{\bsnm{Schuster}, \binits{D.}},
\bauthor{\bsnm{Sears}, \binits{A.}},
\bauthor{\bsnm{Ginossar}, \binits{E.}},
\bauthor{\bsnm{DiCarlo}, \binits{L.}},
\bauthor{\bsnm{Frunzio}, \binits{L.}},
\bauthor{\bsnm{Morton}, \binits{J.}},
\bauthor{\bsnm{Wu}, \binits{H.}},
\bauthor{\bsnm{Briggs}, \binits{G.}},
\bauthor{\bsnm{Buckley}, \binits{B.}},
\bauthor{\bsnm{Awschalom}, \binits{D.}}, \betal:
\batitle{High-cooperativity coupling of electron-spin ensembles to
  superconducting cavities}.
\bjtitle{Physical Review Letters}
\bvolume{105}(\bissue{14}),
\bfpage{140501}
(\byear{2010}).
\doiurl{10.1103/PhysRevLett.105.140501}
\end{barticle}
\endbibitem

\bibitem{paper_amsuss2011cavity}
\begin{barticle}
\bauthor{\bsnm{Ams{\"u}ss}, \binits{R.}},
\bauthor{\bsnm{Koller}, \binits{C.}},
\bauthor{\bsnm{N{\"o}bauer}, \binits{T.}},
\bauthor{\bsnm{Putz}, \binits{S.}},
\bauthor{\bsnm{Rotter}, \binits{S.}},
\bauthor{\bsnm{Sandner}, \binits{K.}},
\bauthor{\bsnm{Schneider}, \binits{S.}},
\bauthor{\bsnm{Schramb{\"o}ck}, \binits{M.}},
\bauthor{\bsnm{Steinhauser}, \binits{G.}},
\bauthor{\bsnm{Ritsch}, \binits{H.}}, \betal:
\batitle{Cavity {QED} with magnetically coupled collective spin states}.
\bjtitle{Physical Review Letters}
\bvolume{107}(\bissue{6}),
\bfpage{060502}
(\byear{2011}).
\doiurl{10.1103/PhysRevLett.107.060502}
\end{barticle}
\endbibitem

\bibitem{paper_astafiev2012coherent}
\begin{barticle}
\bauthor{\bsnm{Astafiev}, \binits{O.}},
\bauthor{\bsnm{Ioffe}, \binits{L.}},
\bauthor{\bsnm{Kafanov}, \binits{S.}},
\bauthor{\bsnm{Pashkin}, \binits{Y.A.}},
\bauthor{\bsnm{Arutyunov}, \binits{K.Y.}},
\bauthor{\bsnm{Shahar}, \binits{D.}},
\bauthor{\bsnm{Cohen}, \binits{O.}},
\bauthor{\bsnm{Tsai}, \binits{J.S.}}:
\batitle{Coherent quantum phase slip}.
\bjtitle{Nature}
\bvolume{484}(\bissue{7394}),
\bfpage{355}--\blpage{358}
(\byear{2012}).
\doiurl{10.1038/nature10930}
\end{barticle}
\endbibitem

\bibitem{paper_peltonen2018hybrid}
\begin{barticle}
\bauthor{\bsnm{Peltonen}, \binits{J.}},
\bauthor{\bsnm{Coumou}, \binits{P.}},
\bauthor{\bsnm{Peng}, \binits{Z.}},
\bauthor{\bsnm{Klapwijk}, \binits{T.}},
\bauthor{\bsnm{Tsai}, \binits{J.}},
\bauthor{\bsnm{Astafiev}, \binits{O.}}:
\batitle{Hybrid rf {SQUID} qubit based on high kinetic inductance}.
\bjtitle{Scientific Reports}
\bvolume{8}(\bissue{1}),
\bfpage{10033}
(\byear{2018}).
\doiurl{10.1038/s41598-018-27154-1}
\end{barticle}
\endbibitem

\bibitem{paper_mooij2005phase}
\begin{barticle}
\bauthor{\bsnm{Mooij}, \binits{J.}},
\bauthor{\bsnm{Harmans}, \binits{C.}}:
\batitle{Phase-slip flux qubits}.
\bjtitle{New Journal of Physics}
\bvolume{7}(\bissue{1}),
\bfpage{219}
(\byear{2005}).
\doiurl{10.1088/1367-2630/7/1/219}
\end{barticle}
\endbibitem

\bibitem{paper_mooij2006superconducting}
\begin{barticle}
\bauthor{\bsnm{Mooij}, \binits{J.}},
\bauthor{\bsnm{Nazarov}, \binits{Y.V.}}:
\batitle{Superconducting nanowires as quantum phase-slip junctions}.
\bjtitle{Nature Physics}
\bvolume{2}(\bissue{3}),
\bfpage{169}--\blpage{172}
(\byear{2006}).
\doiurl{10.1038/nphys234}
\end{barticle}
\endbibitem

\bibitem{paper_schuster2010proposal}
\begin{barticle}
\bauthor{\bsnm{Schuster}, \binits{D.}},
\bauthor{\bsnm{Fragner}, \binits{A.}},
\bauthor{\bsnm{Dykman}, \binits{M.}},
\bauthor{\bsnm{Lyon}, \binits{S.}},
\bauthor{\bsnm{Schoelkopf}, \binits{R.}}:
\batitle{Proposal for manipulating and detecting spin and orbital states of
  trapped electrons on helium using cavity quantum electrodynamics}.
\bjtitle{Physical Review Letters}
\bvolume{105}(\bissue{4}),
\bfpage{040503}
(\byear{2010}).
\doiurl{10.1103/PhysRevLett.105.040503}
\end{barticle}
\endbibitem

\bibitem{paper_bushev2011trapped}
\begin{barticle}
\bauthor{\bsnm{Bushev}, \binits{P.}},
\bauthor{\bsnm{Bothner}, \binits{D.}},
\bauthor{\bsnm{Nagel}, \binits{J.}},
\bauthor{\bsnm{Kemmler}, \binits{M.}},
\bauthor{\bsnm{Konovalenko}, \binits{K.}},
\bauthor{\bsnm{Loerincz}, \binits{A.}},
\bauthor{\bsnm{Ilin}, \binits{K.}},
\bauthor{\bsnm{Siegel}, \binits{M.}},
\bauthor{\bsnm{Koelle}, \binits{D.}},
\bauthor{\bsnm{Kleiner}, \binits{R.}}, \betal:
\batitle{Trapped electron coupled to superconducting devices}.
\bjtitle{The European Physical Journal D}
\bvolume{63},
\bfpage{9}--\blpage{16}
(\byear{2011}).
\doiurl{10.1140/epjd/e2011-10517-6}
\end{barticle}
\endbibitem

\bibitem{Song_resonator_field2009}
\begin{barticle}
\bauthor{\bsnm{Song}, \binits{C.}},
\bauthor{\bsnm{Heitmann}, \binits{T.W.}},
\bauthor{\bsnm{DeFeo}, \binits{M.P.}},
\bauthor{\bsnm{Yu}, \binits{K.}},
\bauthor{\bsnm{McDermott}, \binits{R.}},
\bauthor{\bsnm{Neeley}, \binits{M.}},
\bauthor{\bsnm{Martinis}, \binits{J.M.}},
\bauthor{\bsnm{Plourde}, \binits{B.L.T.}}:
\batitle{Microwave response of vortices in superconducting thin films of {Re}
  and {Al}}.
\bjtitle{Phys. Rev. B}
\bvolume{79},
\bfpage{174512}
(\byear{2009}).
\doiurl{10.1103/PhysRevB.79.174512}
\end{barticle}
\endbibitem

\bibitem{paper_bothner2011improving}
\begin{barticle}
\bauthor{\bsnm{Bothner}, \binits{D.}},
\bauthor{\bsnm{Gaber}, \binits{T.}},
\bauthor{\bsnm{Kemmler}, \binits{M.}},
\bauthor{\bsnm{Koelle}, \binits{D.}},
\bauthor{\bsnm{Kleiner}, \binits{R.}}:
\batitle{Improving the performance of superconducting microwave resonators in
  magnetic fields}.
\bjtitle{Applied Physics Letters}
\bvolume{98}(\bissue{10}),
\bfpage{102504}
(\byear{2011}).
\doiurl{10.1063/1.3560480}
\end{barticle}
\endbibitem

\bibitem{paper_bothner2012reducing}
\begin{barticle}
\bauthor{\bsnm{Bothner}, \binits{D.}},
\bauthor{\bsnm{Clauss}, \binits{C.}},
\bauthor{\bsnm{Koroknay}, \binits{E.}},
\bauthor{\bsnm{Kemmler}, \binits{M.}},
\bauthor{\bsnm{Gaber}, \binits{T.}},
\bauthor{\bsnm{Jetter}, \binits{M.}},
\bauthor{\bsnm{Scheffler}, \binits{M.}},
\bauthor{\bsnm{Michler}, \binits{P.}},
\bauthor{\bsnm{Dressel}, \binits{M.}},
\bauthor{\bsnm{Koelle}, \binits{D.}}, \betal:
\batitle{Reducing vortex losses in superconducting microwave resonators with
  microsphere patterned antidot arrays}.
\bjtitle{Applied Physics Letters}
\bvolume{100}(\bissue{1}),
\bfpage{012601}
(\byear{2012}).
\doiurl{10.1063/1.3673869}
\end{barticle}
\endbibitem

\bibitem{paper_bothner2012magnetic}
\begin{barticle}
\bauthor{\bsnm{Bothner}, \binits{D.}},
\bauthor{\bsnm{Gaber}, \binits{T.}},
\bauthor{\bsnm{Kemmler}, \binits{M.}},
\bauthor{\bsnm{Koelle}, \binits{D.}},
\bauthor{\bsnm{Kleiner}, \binits{R.}},
\bauthor{\bsnm{W{\"u}nsch}, \binits{S.}},
\bauthor{\bsnm{Siegel}, \binits{M.}}:
\batitle{Magnetic hysteresis effects in superconducting coplanar microwave
  resonators}.
\bjtitle{Physical Review B}
\bvolume{86}(\bissue{1}),
\bfpage{014517}
(\byear{2012}).
\doiurl{10.1103/PhysRevB.86.014517}
\end{barticle}
\endbibitem

\bibitem{paper_chiaro2016dielectric}
\begin{barticle}
\bauthor{\bsnm{Chiaro}, \binits{B.}},
\bauthor{\bsnm{Megrant}, \binits{A.}},
\bauthor{\bsnm{Dunsworth}, \binits{A.}},
\bauthor{\bsnm{Chen}, \binits{Z.}},
\bauthor{\bsnm{Barends}, \binits{R.}},
\bauthor{\bsnm{Campbell}, \binits{B.}},
\bauthor{\bsnm{Chen}, \binits{Y.}},
\bauthor{\bsnm{Fowler}, \binits{A.}},
\bauthor{\bsnm{Hoi}, \binits{I.}},
\bauthor{\bsnm{Jeffrey}, \binits{E.}}, \betal:
\batitle{Dielectric surface loss in superconducting resonators with
  flux-trapping holes}.
\bjtitle{Superconductor Science and Technology}
\bvolume{29}(\bissue{10}),
\bfpage{104006}
(\byear{2016}).
\doiurl{10.1088/0953-2048/29/10/104006}
\end{barticle}
\endbibitem

\bibitem{paper_kroll2019magnetic}
\begin{barticle}
\bauthor{\bsnm{Kroll}, \binits{J.G.}},
\bauthor{\bsnm{Borsoi}, \binits{F.}},
\bauthor{\bsnm{Van Der~Enden}, \binits{K.}},
\bauthor{\bsnm{Uilhoorn}, \binits{W.}},
\bauthor{\bsnm{De~Jong}, \binits{D.}},
\bauthor{\bsnm{Quintero-P{\'e}rez}, \binits{M.}},
\bauthor{\bsnm{Van~Woerkom}, \binits{D.}},
\bauthor{\bsnm{Bruno}, \binits{A.}},
\bauthor{\bsnm{Plissard}, \binits{S.}},
\bauthor{\bsnm{Car}, \binits{D.}}, \betal:
\batitle{Magnetic-field-resilient superconducting coplanar-waveguide resonators
  for hybrid circuit quantum electrodynamics experiments}.
\bjtitle{Physical Review Applied}
\bvolume{11}(\bissue{6}),
\bfpage{064053}
(\byear{2019}).
\doiurl{10.1103/PhysRevApplied.11.064053}
\end{barticle}
\endbibitem

\bibitem{Raes2012_singlevortexmotion}
\begin{barticle}
\bauthor{\bsnm{Raes}, \binits{B.}},
\bauthor{\bparticle{Van~de} \bsnm{Vondel}, \binits{J.}},
\bauthor{\bsnm{Silhanek}, \binits{A.V.}},
\bauthor{\bparticle{de} \bsnm{Souza~Silva}, \binits{C.C.}},
\bauthor{\bsnm{Gutierrez}, \binits{J.}},
\bauthor{\bsnm{Kramer}, \binits{R.B.G.}},
\bauthor{\bsnm{Moshchalkov}, \binits{V.V.}}:
\batitle{Local mapping of dissipative vortex motion}.
\bjtitle{Phys. Rev. B}
\bvolume{86},
\bfpage{064522}
(\byear{2012}).
\doiurl{10.1103/PhysRevB.86.064522}
\end{barticle}
\endbibitem

\bibitem{Nsanzineza_impact_singlevortex2014}
\begin{barticle}
\bauthor{\bsnm{Nsanzineza}, \binits{I.}},
\bauthor{\bsnm{Plourde}, \binits{B.L.T.}}:
\batitle{Trapping a single vortex and reducing quasiparticles in a
  superconducting resonator}.
\bjtitle{Phys. Rev. Lett.}
\bvolume{113},
\bfpage{117002}
(\byear{2014}).
\doiurl{10.1103/PhysRevLett.113.117002}
\end{barticle}
\endbibitem

\bibitem{paper_song2009reducing}
\begin{barticle}
\bauthor{\bsnm{Song}, \binits{C.}},
\bauthor{\bsnm{DeFeo}, \binits{M.P.}},
\bauthor{\bsnm{Yu}, \binits{K.}},
\bauthor{\bsnm{Plourde}, \binits{B.L.}}:
\batitle{Reducing microwave loss in superconducting resonators due to trapped
  vortices}.
\bjtitle{Applied Physics Letters}
\bvolume{95}(\bissue{23}),
\bfpage{232501}
(\byear{2009}).
\doiurl{10.1063/1.3271523}
\end{barticle}
\endbibitem

\bibitem{paper_bothner2017improving}
\begin{barticle}
\bauthor{\bsnm{Bothner}, \binits{D.}},
\bauthor{\bsnm{Wiedmaier}, \binits{D.}},
\bauthor{\bsnm{Ferdinand}, \binits{B.}},
\bauthor{\bsnm{Kleiner}, \binits{R.}},
\bauthor{\bsnm{Koelle}, \binits{D.}}:
\batitle{Improving superconducting resonators in magnetic fields by reduced
  field focussing and engineered flux screening}.
\bjtitle{Physical Review Applied}
\bvolume{8}(\bissue{3}),
\bfpage{034025}
(\byear{2017}).
\doiurl{10.1103/PhysRevApplied.8.034025}
\end{barticle}
\endbibitem

\bibitem{paper_graaf2012magnetic}
\begin{barticle}
\bauthor{\bsnm{Graaf}, \binits{S.d.}},
\bauthor{\bsnm{Danilov}, \binits{A.}},
\bauthor{\bsnm{Adamyan}, \binits{A.}},
\bauthor{\bsnm{Bauch}, \binits{T.}},
\bauthor{\bsnm{Kubatkin}, \binits{S.}}:
\batitle{Magnetic field resilient superconducting fractal resonators for
  coupling to free spins}.
\bjtitle{Journal of Applied Physics}
\bvolume{112}(\bissue{12}),
\bfpage{123905}
(\byear{2012}).
\doiurl{10.1063/1.4769208}
\end{barticle}
\endbibitem

\bibitem{paper_de2014galvanically}
\begin{barticle}
\bauthor{\bparticle{de} \bsnm{Graaf}, \binits{S.E.}},
\bauthor{\bsnm{Davidovikj}, \binits{D.}},
\bauthor{\bsnm{Adamyan}, \binits{A.}},
\bauthor{\bsnm{Kubatkin}, \binits{S.}},
\bauthor{\bsnm{Danilov}, \binits{A.}}:
\batitle{Galvanically split superconducting microwave resonators for
  introducing internal voltage bias}.
\bjtitle{Applied Physics Letters}
\bvolume{104}(\bissue{5}),
\bfpage{052601}
(\byear{2014}).
\doiurl{10.1063/1.4863681}
\end{barticle}
\endbibitem

\bibitem{paper_lange2017high}
\begin{barticle}
\bauthor{\bsnm{Lange}, \binits{M.}},
\bauthor{\bsnm{Gu{\'e}non}, \binits{S.}},
\bauthor{\bsnm{Lever}, \binits{F.}},
\bauthor{\bsnm{Kleiner}, \binits{R.}},
\bauthor{\bsnm{Koelle}, \binits{D.}}:
\batitle{A high-resolution combined scanning laser and widefield polarizing
  microscope for imaging at temperatures from 4 {K} to 300 {K}}.
\bjtitle{Review of Scientific Instruments}
\bvolume{88}(\bissue{12}),
\bfpage{123705}
(\byear{2017}).
\doiurl{10.1063/1.5009529}
\end{barticle}
\endbibitem

\bibitem{paper_ghigo2007evidence}
\begin{barticle}
\bauthor{\bsnm{Ghigo}, \binits{G.}},
\bauthor{\bsnm{Laviano}, \binits{F.}},
\bauthor{\bsnm{Gozzelino}, \binits{L.}},
\bauthor{\bsnm{Gerbaldo}, \binits{R.}},
\bauthor{\bsnm{Mezzetti}, \binits{E.}},
\bauthor{\bsnm{Monticone}, \binits{E.}},
\bauthor{\bsnm{Portesi}, \binits{C.}}:
\batitle{Evidence of rf-driven dendritic vortex avalanches in {MgB}$_2$
  microwave resonators}.
\bjtitle{Journal of Applied Physics}
\bvolume{102}(\bissue{11}),
\bfpage{113901}
(\byear{2007}).
\doiurl{10.1063/1.2816257}
\end{barticle}
\endbibitem

\bibitem{paper_doyle2008lumped}
\begin{barticle}
\bauthor{\bsnm{Doyle}, \binits{S.}},
\bauthor{\bsnm{Mauskopf}, \binits{P.}},
\bauthor{\bsnm{Naylon}, \binits{J.}},
\bauthor{\bsnm{Porch}, \binits{A.}},
\bauthor{\bsnm{Duncombe}, \binits{C.}}:
\batitle{Lumped element kinetic inductance detectors}.
\bjtitle{Journal of Low Temperature Physics}
\bvolume{151},
\bfpage{530}--\blpage{536}
(\byear{2008}).
\doiurl{10.1007/s10909-007-9685-2}
\end{barticle}
\endbibitem

\bibitem{paper_grabovskij2008situ}
\begin{barticle}
\bauthor{\bsnm{Grabovskij}, \binits{G.}},
\bauthor{\bsnm{Swenson}, \binits{L.}},
\bauthor{\bsnm{Buisson}, \binits{O.}},
\bauthor{\bsnm{Hoffmann}, \binits{C.}},
\bauthor{\bsnm{Monfardini}, \binits{A.}},
\bauthor{\bsnm{Vill{\'e}gier}, \binits{J.-C.}}:
\batitle{In situ measurement of the permittivity of helium using microwave
  {NbN} resonators}.
\bjtitle{Applied Physics Letters}
\bvolume{93}(\bissue{13}),
\bfpage{134102}
(\byear{2008}).
\doiurl{10.1063/1.2996263}
\end{barticle}
\endbibitem

\bibitem{paper_khalil2012analysis}
\begin{barticle}
\bauthor{\bsnm{Khalil}, \binits{M.S.}},
\bauthor{\bsnm{Stoutimore}, \binits{M.}},
\bauthor{\bsnm{Wellstood}, \binits{F.}},
\bauthor{\bsnm{Osborn}, \binits{K.}}:
\batitle{An analysis method for asymmetric resonator transmission applied to
  superconducting devices}.
\bjtitle{Journal of Applied Physics}
\bvolume{111}(\bissue{5}),
\bfpage{054510}
(\byear{2012}).
\doiurl{10.1063/1.3692073}
\end{barticle}
\endbibitem

\bibitem{paper_probst2015efficient}
\begin{barticle}
\bauthor{\bsnm{Probst}, \binits{S.}},
\bauthor{\bsnm{Song}, \binits{F.}},
\bauthor{\bsnm{Bushev}, \binits{P.A.}},
\bauthor{\bsnm{Ustinov}, \binits{A.V.}},
\bauthor{\bsnm{Weides}, \binits{M.}}:
\batitle{Efficient and robust analysis of complex scattering data under noise
  in microwave resonators}.
\bjtitle{Review of Scientific Instruments}
\bvolume{86}(\bissue{2}),
\bfpage{024706}
(\byear{2015}).
\doiurl{10.1063/1.4907935}
\end{barticle}
\endbibitem

\bibitem{Book_pozar2011microwave}
\begin{bbook}
\bauthor{\bsnm{Pozar}, \binits{D.M.}}:
\bbtitle{Microwave Engineering}.
\bpublisher{John wiley \& sons},
\blocation{Hoboken, New Jersey}
(\byear{2011})
\end{bbook}
\endbibitem

\bibitem{Book_tinkham2004introduction}
\begin{bbook}
\bauthor{\bsnm{Tinkham}, \binits{M.}}:
\bbtitle{Introduction to Superconductivity},
\bedition{2}nd edn.
\bpublisher{Dover Publications},
\blocation{Mineola, NY}
(\byear{2004})
\end{bbook}
\endbibitem

\bibitem{paper_annunziata2010tunable}
\begin{barticle}
\bauthor{\bsnm{Annunziata}, \binits{A.J.}},
\bauthor{\bsnm{Santavicca}, \binits{D.F.}},
\bauthor{\bsnm{Frunzio}, \binits{L.}},
\bauthor{\bsnm{Catelani}, \binits{G.}},
\bauthor{\bsnm{Rooks}, \binits{M.J.}},
\bauthor{\bsnm{Frydman}, \binits{A.}},
\bauthor{\bsnm{Prober}, \binits{D.E.}}:
\batitle{Tunable superconducting nanoinductors}.
\bjtitle{Nanotechnology}
\bvolume{21}(\bissue{44}),
\bfpage{445202}
(\byear{2010}).
\doiurl{10.1088/0957-4484/21/44/445202}
\end{barticle}
\endbibitem

\bibitem{paper_bonura2006magnetic}
\begin{barticle}
\bauthor{\bsnm{Bonura}, \binits{M.}},
\bauthor{\bsnm{Agliolo~Gallitto}, \binits{A.}},
\bauthor{\bsnm{Li~Vigni}, \binits{M.}}:
\batitle{Magnetic hysteresis in the microwave surface resistance of {Nb}
  samples in the critical state}.
\bjtitle{The European Physical Journal B-Condensed Matter and Complex Systems}
\bvolume{53},
\bfpage{315}--\blpage{322}
(\byear{2006}).
\doiurl{10.1140/epjb/e2006-00381-8}
\end{barticle}
\endbibitem

\bibitem{paper_yu2021magnetic}
\begin{barticle}
\bauthor{\bsnm{Yu}, \binits{C.X.}},
\bauthor{\bsnm{Zihlmann}, \binits{S.}},
\bauthor{\bsnm{Troncoso~Fern{\'a}ndez-Bada}, \binits{G.}},
\bauthor{\bsnm{Thomassin}, \binits{J.-L.}},
\bauthor{\bsnm{Gustavo}, \binits{F.}},
\bauthor{\bsnm{Dumur}, \binits{{\'E}.}},
\bauthor{\bsnm{Maurand}, \binits{R.}}:
\batitle{Magnetic field resilient high kinetic inductance superconducting
  niobium nitride coplanar waveguide resonators}.
\bjtitle{Applied Physics Letters}
\bvolume{118}(\bissue{5}),
\bfpage{054001}
(\byear{2021}).
\doiurl{10.1063/5.0039945}
\end{barticle}
\endbibitem

\bibitem{paper_lahl2003nonlinear}
\begin{barticle}
\bauthor{\bsnm{Lahl}, \binits{P.}},
\bauthor{\bsnm{Wordenweber}, \binits{R.}}:
\batitle{Nonlinear microwave properties of {HTS} thin film coplanar devices}.
\bjtitle{IEEE transactions on applied superconductivity}
\bvolume{13}(\bissue{2}),
\bfpage{2917}--\blpage{2920}
(\byear{2003}).
\doiurl{10.1109/TASC.2003.812046}
\end{barticle}
\endbibitem

\bibitem{paper_ghigo2010mechanisms}
\begin{barticle}
\bauthor{\bsnm{Ghigo}, \binits{G.}},
\bauthor{\bsnm{Gerbaldo}, \binits{R.}},
\bauthor{\bsnm{Gozzelino}, \binits{L.}},
\bauthor{\bsnm{Laviano}, \binits{F.}},
\bauthor{\bsnm{Minetti}, \binits{B.}},
\bauthor{\bsnm{Monticone}, \binits{E.}},
\bauthor{\bsnm{Portesi}, \binits{C.}},
\bauthor{\bsnm{Mezzetti}, \binits{E.}}:
\batitle{Mechanisms limiting the performance of {MgB}$_2$ polycrystalline thin
  film microwave resonators}.
\bjtitle{IEEE transactions on applied superconductivity}
\bvolume{21}(\bissue{3}),
\bfpage{579}--\blpage{582}
(\byear{2010}).
\doiurl{10.1109/TASC.2010.2093107}
\end{barticle}
\endbibitem

\bibitem{paper_borisov2020superconducting}
\begin{barticle}
\bauthor{\bsnm{Borisov}, \binits{K.}},
\bauthor{\bsnm{Rieger}, \binits{D.}},
\bauthor{\bsnm{Winkel}, \binits{P.}},
\bauthor{\bsnm{Henriques}, \binits{F.}},
\bauthor{\bsnm{Valenti}, \binits{F.}},
\bauthor{\bsnm{Ionita}, \binits{A.}},
\bauthor{\bsnm{Wessbecher}, \binits{M.}},
\bauthor{\bsnm{Spiecker}, \binits{M.}},
\bauthor{\bsnm{Gusenkova}, \binits{D.}},
\bauthor{\bsnm{Pop}, \binits{I.}}, \betal:
\batitle{Superconducting granular aluminum resonators resilient to magnetic
  fields up to 1 tesla}.
\bjtitle{Applied Physics Letters}
\bvolume{117}(\bissue{12}),
\bfpage{120502}
(\byear{2020})
\end{barticle}
\endbibitem

\bibitem{paper_olsen2007avalanches}
\begin{barticle}
\bauthor{\bsnm{Olsen}, \binits{{\AA}.A.F.}},
\bauthor{\bsnm{Johansen}, \binits{T.H.}},
\bauthor{\bsnm{Shantsev}, \binits{D.}},
\bauthor{\bsnm{Choi}, \binits{E.-M.}},
\bauthor{\bsnm{Lee}, \binits{H.-S.}},
\bauthor{\bsnm{Kim}, \binits{H.J.}},
\bauthor{\bsnm{Lee}, \binits{S.-I.}}:
\batitle{Avalanches injecting flux into the central hole of a superconducting
  {MgB}$_2$ ring}.
\bjtitle{Physical Review B}
\bvolume{76}(\bissue{2}),
\bfpage{024510}
(\byear{2007}).
\doiurl{10.1103/PhysRevB.76.024510}
\end{barticle}
\endbibitem

\bibitem{paper_shvartzberg2019quasiperiodic}
\begin{barticle}
\bauthor{\bsnm{Shvartzberg}, \binits{J.}},
\bauthor{\bsnm{Shaulov}, \binits{A.}},
\bauthor{\bsnm{Yeshurun}, \binits{Y.}}:
\batitle{Quasiperiodic magnetic flux avalanches in doubly connected
  superconductors}.
\bjtitle{Physical Review B}
\bvolume{100}(\bissue{18}),
\bfpage{184506}
(\byear{2019}).
\doiurl{10.1103/PhysRevB.100.184506}
\end{barticle}
\endbibitem

\bibitem{paper_jiang2020selective}
\begin{barticle}
\bauthor{\bsnm{Jiang}, \binits{L.}},
\bauthor{\bsnm{Xue}, \binits{C.}},
\bauthor{\bsnm{Burger}, \binits{L.}},
\bauthor{\bsnm{Vanderheyden}, \binits{B.}},
\bauthor{\bsnm{Silhanek}, \binits{A.}},
\bauthor{\bsnm{Zhou}, \binits{Y.-H.}}:
\batitle{Selective triggering of magnetic flux avalanches by an edge
  indentation}.
\bjtitle{Physical Review B}
\bvolume{101}(\bissue{22}),
\bfpage{224505}
(\byear{2020}).
\doiurl{10.1103/PhysRevB.101.224505}
\end{barticle}
\endbibitem

\bibitem{RevModPhys_criticalstate}
\begin{barticle}
\bauthor{\bsnm{Mints}, \binits{R.G.}},
\bauthor{\bsnm{Rakhmanov}, \binits{A.L.}}:
\batitle{Critical state stability in type-ii superconductors and
  superconducting-normal-metal composites}.
\bjtitle{Rev. Mod. Phys.}
\bvolume{53},
\bfpage{551}--\blpage{592}
(\byear{1981}).
\doiurl{10.1103/RevModPhys.53.551}
\end{barticle}
\endbibitem

\bibitem{paper_johansen2007reproducible}
\begin{barticle}
\bauthor{\bsnm{Johansen}, \binits{T.}},
\bauthor{\bsnm{Shantsev}, \binits{D.}},
\bauthor{\bsnm{Olsen}, \binits{{\AA}.A.}},
\bauthor{\bsnm{Roussel}, \binits{M.}},
\bauthor{\bsnm{Pan}, \binits{A.}},
\bauthor{\bsnm{Dou}, \binits{S.}}:
\batitle{Reproducible nucleation sites for flux dendrites in {MgB}$_2$}.
\bjtitle{Surface science}
\bvolume{601}(\bissue{24}),
\bfpage{5712}--\blpage{5714}
(\byear{2007}).
\doiurl{10.1016/j.susc.2007.06.068}
\end{barticle}
\endbibitem

\bibitem{paper_qureishy2017dendritic}
\begin{barticle}
\bauthor{\bsnm{Qureishy}, \binits{T.}},
\bauthor{\bsnm{Laliena}, \binits{C.}},
\bauthor{\bsnm{Mart{\'\i}nez}, \binits{E.}},
\bauthor{\bsnm{Qviller}, \binits{A.J.}},
\bauthor{\bsnm{Vestg{\aa}rden}, \binits{J.I.}},
\bauthor{\bsnm{Johansen}, \binits{T.H.}},
\bauthor{\bsnm{Navarro}, \binits{R.}},
\bauthor{\bsnm{Mikheenko}, \binits{P.}}:
\batitle{Dendritic flux avalanches in a superconducting {MgB}$_2$ tape}.
\bjtitle{Superconductor Science and Technology}
\bvolume{30}(\bissue{12}),
\bfpage{125005}
(\byear{2017}).
\doiurl{10.1088/1361-6668/aa9244}
\end{barticle}
\endbibitem

\bibitem{paper_brisbois2016magnetic}
\begin{barticle}
\bauthor{\bsnm{Brisbois}, \binits{J.}},
\bauthor{\bsnm{Adami}, \binits{O.-A.}},
\bauthor{\bsnm{Avila}, \binits{J.}},
\bauthor{\bsnm{Motta}, \binits{M.}},
\bauthor{\bsnm{Ortiz}, \binits{W.A.}},
\bauthor{\bsnm{Nguyen}, \binits{N.D.}},
\bauthor{\bsnm{Vanderbemden}, \binits{P.}},
\bauthor{\bsnm{Vanderheyden}, \binits{B.}},
\bauthor{\bsnm{Kramer}, \binits{R.B.}},
\bauthor{\bsnm{Silhanek}, \binits{A.}}:
\batitle{Magnetic flux penetration in {Nb} superconducting films with
  lithographically defined microindentations}.
\bjtitle{Physical Review B}
\bvolume{93}(\bissue{5}),
\bfpage{054521}
(\byear{2016}).
\doiurl{10.1103/PhysRevB.93.054521}
\end{barticle}
\endbibitem

\bibitem{Brandt1993_edgeH}
\begin{barticle}
\bauthor{\bsnm{Brandt}, \binits{E.H.}},
\bauthor{\bsnm{Indenbom}, \binits{M.}}:
\batitle{Type-ii-superconductor strip with current in a perpendicular magnetic
  field}.
\bjtitle{Phys. Rev. B}
\bvolume{48},
\bfpage{12893}--\blpage{12906}
(\byear{1993}).
\doiurl{10.1103/PhysRevB.48.12893}
\end{barticle}
\endbibitem

\bibitem{Zeldov1994_edgeH}
\begin{barticle}
\bauthor{\bsnm{Zeldov}, \binits{E.}},
\bauthor{\bsnm{Clem}, \binits{J.R.}},
\bauthor{\bsnm{McElfresh}, \binits{M.}},
\bauthor{\bsnm{Darwin}, \binits{M.}}:
\batitle{Magnetization and transport currents in thin superconducting films}.
\bjtitle{Phys. Rev. B}
\bvolume{49},
\bfpage{9802}--\blpage{9822}
(\byear{1994}).
\doiurl{10.1103/PhysRevB.49.9802}
\end{barticle}
\endbibitem

\bibitem{McDonald1996_edgeH}
\begin{barticle}
\bauthor{\bsnm{McDonald}, \binits{J.}},
\bauthor{\bsnm{Clem}, \binits{J.R.}}:
\batitle{Theory of flux penetration into thin films with field-dependent
  critical current}.
\bjtitle{Phys. Rev. B}
\bvolume{53},
\bfpage{8643}--\blpage{8650}
(\byear{1996}).
\doiurl{10.1103/PhysRevB.53.8643}
\end{barticle}
\endbibitem

\bibitem{paper_motta2011visualizing}
\begin{barticle}
\bauthor{\bsnm{Motta}, \binits{M.}},
\bauthor{\bsnm{Colauto}, \binits{F.}},
\bauthor{\bsnm{Zadorosny}, \binits{R.}},
\bauthor{\bsnm{Johansen}, \binits{T.}},
\bauthor{\bsnm{Dinner}, \binits{R.}},
\bauthor{\bsnm{Blamire}, \binits{M.}},
\bauthor{\bsnm{Ataklti}, \binits{G.}},
\bauthor{\bsnm{Moshchalkov}, \binits{V.}},
\bauthor{\bsnm{Silhanek}, \binits{A.}},
\bauthor{\bsnm{Ortiz}, \binits{W.}}:
\batitle{Visualizing the ac magnetic susceptibility of superconducting films
  via magneto-optical imaging}.
\bjtitle{Physical Review B}
\bvolume{84}(\bissue{21}),
\bfpage{214529}
(\byear{2011}).
\doiurl{10.1103/PhysRevB.84.214529}
\end{barticle}
\endbibitem

\bibitem{paper_gurevich2000nonlinear}
\begin{barticle}
\bauthor{\bsnm{Gurevich}, \binits{A.}},
\bauthor{\bsnm{Friesen}, \binits{M.}}:
\batitle{Nonlinear transport current flow in superconductors with planar
  obstacles}.
\bjtitle{Physical Review B}
\bvolume{62}(\bissue{6}),
\bfpage{4004}
(\byear{2000}).
\doiurl{10.1103/PhysRevB.62.4004}
\end{barticle}
\endbibitem

\bibitem{paper_zeldov1994magnetization}
\begin{barticle}
\bauthor{\bsnm{Zeldov}, \binits{E.}},
\bauthor{\bsnm{Clem}, \binits{J.R.}},
\bauthor{\bsnm{McElfresh}, \binits{M.}},
\bauthor{\bsnm{Darwin}, \binits{M.}}:
\batitle{Magnetization and transport currents in thin superconducting films}.
\bjtitle{Physical Review B}
\bvolume{49}(\bissue{14}),
\bfpage{9802}
(\byear{1994}).
\doiurl{10.1103/PhysRevB.49.9802}
\end{barticle}
\endbibitem

\bibitem{paper_denisov2006onset}
\begin{barticle}
\bauthor{\bsnm{Denisov}, \binits{D.}},
\bauthor{\bsnm{Shantsev}, \binits{D.}},
\bauthor{\bsnm{Galperin}, \binits{Y.}},
\bauthor{\bsnm{Choi}, \binits{E.-M.}},
\bauthor{\bsnm{Lee}, \binits{H.-S.}},
\bauthor{\bsnm{Lee}, \binits{S.-I.}},
\bauthor{\bsnm{Bobyl}, \binits{A.}},
\bauthor{\bsnm{Goa}, \binits{P.}},
\bauthor{\bsnm{Olsen}, \binits{A.}},
\bauthor{\bsnm{Johansen}, \binits{T.}}:
\batitle{Onset of dendritic flux avalanches in superconducting films}.
\bjtitle{Physical Review Letters}
\bvolume{97}(\bissue{7}),
\bfpage{077002}
(\byear{2006}).
\doiurl{10.1103/PhysRevLett.97.077002}
\end{barticle}
\endbibitem

\bibitem{paper_samkharadze2016high}
\begin{barticle}
\bauthor{\bsnm{Samkharadze}, \binits{N.}},
\bauthor{\bsnm{Bruno}, \binits{A.}},
\bauthor{\bsnm{Scarlino}, \binits{P.}},
\bauthor{\bsnm{Zheng}, \binits{G.}},
\bauthor{\bsnm{DiVincenzo}, \binits{D.}},
\bauthor{\bsnm{DiCarlo}, \binits{L.}},
\bauthor{\bsnm{Vandersypen}, \binits{L.}}:
\batitle{High-kinetic-inductance superconducting nanowire resonators for
  circuit {QED} in a magnetic field}.
\bjtitle{Physical Review Applied}
\bvolume{5}(\bissue{4}),
\bfpage{044004}
(\byear{2016}).
\doiurl{10.1103/PhysRevApplied.5.044004}
\end{barticle}
\endbibitem

\bibitem{paper_cerbu2013vortex}
\begin{barticle}
\bauthor{\bsnm{Cerbu}, \binits{D.}},
\bauthor{\bsnm{Gladilin}, \binits{V.}},
\bauthor{\bsnm{Cuppens}, \binits{J.}},
\bauthor{\bsnm{Fritzsche}, \binits{J.}},
\bauthor{\bsnm{Tempere}, \binits{J.}},
\bauthor{\bsnm{Devreese}, \binits{J.}},
\bauthor{\bsnm{Moshchalkov}, \binits{V.}},
\bauthor{\bsnm{Silhanek}, \binits{A.}},
\bauthor{\bparticle{Van~de} \bsnm{Vondel}, \binits{J.}}:
\batitle{Vortex ratchet induced by controlled edge roughness}.
\bjtitle{New Journal of Physics}
\bvolume{15}(\bissue{6}),
\bfpage{063022}
(\byear{2013}).
\doiurl{10.1088/1367-2630/15/6/063022}
\end{barticle}
\endbibitem

\bibitem{paper_menghini2005dendritic}
\begin{barticle}
\bauthor{\bsnm{Menghini}, \binits{M.}},
\bauthor{\bsnm{Wijngaarden}, \binits{R.}},
\bauthor{\bsnm{Silhanek}, \binits{A.}},
\bauthor{\bsnm{Raedts}, \binits{S.}},
\bauthor{\bsnm{Moshchalkov}, \binits{V.}}:
\batitle{Dendritic flux penetration in {Pb} films with a periodic array of
  antidots}.
\bjtitle{Physical Review B}
\bvolume{71}(\bissue{10}),
\bfpage{104506}
(\byear{2005}).
\doiurl{10.1103/PhysRevB.71.104506}
\end{barticle}
\endbibitem

\bibitem{paper_motta2014controllable}
\begin{barticle}
\bauthor{\bsnm{Motta}, \binits{M.}},
\bauthor{\bsnm{Colauto}, \binits{F.}},
\bauthor{\bsnm{Vestg\aa{}rden}, \binits{J.I.}},
\bauthor{\bsnm{Fritzsche}, \binits{J.}},
\bauthor{\bsnm{Timmermans}, \binits{M.}},
\bauthor{\bsnm{Cuppens}, \binits{J.}},
\bauthor{\bsnm{Attanasio}, \binits{C.}},
\bauthor{\bsnm{Cirillo}, \binits{C.}},
\bauthor{\bsnm{Moshchalkov}, \binits{V.V.}},
\bauthor{\bparticle{Van~de} \bsnm{Vondel}, \binits{J.}},
\bauthor{\bsnm{Johansen}, \binits{T.H.}},
\bauthor{\bsnm{Ortiz}, \binits{W.A.}},
\bauthor{\bsnm{Silhanek}, \binits{A.V.}}:
\batitle{Controllable morphology of flux avalanches in microstructured
  superconductors}.
\bjtitle{Phys. Rev. B}
\bvolume{89},
\bfpage{134508}
(\byear{2014}).
\doiurl{10.1103/PhysRevB.89.134508}
\end{barticle}
\endbibitem

\bibitem{paper_ghigo2004microwave}
\begin{barticle}
\bauthor{\bsnm{Ghigo}, \binits{G.}},
\bauthor{\bsnm{Botta}, \binits{D.}},
\bauthor{\bsnm{Chiodoni}, \binits{A.}},
\bauthor{\bsnm{Gerbaldo}, \binits{R.}},
\bauthor{\bsnm{Gozzelino}, \binits{L.}},
\bauthor{\bsnm{Laviano}, \binits{F.}},
\bauthor{\bsnm{Minetti}, \binits{B.}},
\bauthor{\bsnm{Mezzetti}, \binits{E.}},
\bauthor{\bsnm{Andreone}, \binits{D.}}:
\batitle{Microwave dissipation in {YBCO} coplanar resonators with uniform and
  non-uniform columnar defect distribution}.
\bjtitle{Superconductor Science and Technology}
\bvolume{17}(\bissue{8}),
\bfpage{977}
(\byear{2004}).
\doiurl{10.1088/0953-2048/17/8/004}
\end{barticle}
\endbibitem

\bibitem{paper_lee1999reducing}
\begin{barticle}
\bauthor{\bsnm{Lee}, \binits{C.-S.}},
\bauthor{\bsnm{Janko}, \binits{B.}},
\bauthor{\bsnm{Derenyi}, \binits{I.}},
\bauthor{\bsnm{Barab{\'a}si}, \binits{A.-L.}}:
\batitle{Reducing vortex density in superconductors using the ‘ratchet
  effect’}.
\bjtitle{Nature}
\bvolume{400}(\bissue{6742}),
\bfpage{337}--\blpage{340}
(\byear{1999})
\end{barticle}
\endbibitem

\bibitem{paper_dobrovolskiy2020upper}
\begin{barticle}
\bauthor{\bsnm{Dobrovolskiy}, \binits{O.}},
\bauthor{\bsnm{Begun}, \binits{E.}},
\bauthor{\bsnm{Bevz}, \binits{V.}},
\bauthor{\bsnm{Sachser}, \binits{R.}},
\bauthor{\bsnm{Huth}, \binits{M.}}:
\batitle{Upper frequency limits for vortex guiding and ratchet effects}.
\bjtitle{Physical Review Applied}
\bvolume{13}(\bissue{2}),
\bfpage{024012}
(\byear{2020}).
\doiurl{10.1103/PhysRevApplied.13.024012}
\end{barticle}
\endbibitem

\bibitem{paper_awad2011flux}
\begin{barticle}
\bauthor{\bsnm{Awad}, \binits{A.}},
\bauthor{\bsnm{Aliev}, \binits{F.G.}},
\bauthor{\bsnm{Ataklti}, \binits{G.}},
\bauthor{\bsnm{Silhanek}, \binits{A.}},
\bauthor{\bsnm{Moshchalkov}, \binits{V.V.}},
\bauthor{\bsnm{Galperin}, \binits{Y.}},
\bauthor{\bsnm{Vinokur}, \binits{V.}}:
\batitle{Flux avalanches triggered by microwave depinning of magnetic vortices
  in {Pb} superconducting films}.
\bjtitle{Physical Review B}
\bvolume{84}(\bissue{22}),
\bfpage{224511}
(\byear{2011}).
\doiurl{10.1103/PhysRevB.84.224511}
\end{barticle}
\endbibitem

\bibitem{Kirby2008}
\begin{botherref}
\oauthor{\bsnm{Kirby}, \binits{K.W.}}:
Processing of Sapphire Surfaces for Semiconductor Device Applications
\end{botherref}
\endbibitem

\bibitem{Burton2015}
\begin{bchapter}
\bauthor{\bsnm{Burton}, \binits{M.C.}},
\bauthor{\bsnm{Beebe}, \binits{M.}},
\bauthor{\bsnm{Lukaszew}, \binits{R.A.}},
\bauthor{\bsnm{Riso}, \binits{J.M.}},
\bauthor{\bsnm{Reece}, \binits{C.E.}},
\bauthor{\bsnm{Valente-Feliciano}, \binits{A.-M.}}:
\bctitle{Superconducting nbn-based multilayer and nbtin thin films for the
  enhancement of srf accelerator cavities}.
(\byear{2015}).
\doiurl{10.18429/JACoW-SRF2015-TUPB037}
\end{bchapter}
\endbibitem

\bibitem{Mahashabde2020}
\begin{barticle}
\bauthor{\bsnm{Mahashabde}, \binits{S.}},
\bauthor{\bsnm{Otto}, \binits{E.}},
\bauthor{\bsnm{Montemurro}, \binits{D.}},
\bauthor{\bparticle{de} \bsnm{Graaf}, \binits{S.}},
\bauthor{\bsnm{Kubatkin}, \binits{S.}},
\bauthor{\bsnm{Danilov}, \binits{A.}}:
\batitle{Fast {Tunable} {High}-\${Q}\$-{Factor} {Superconducting} {Microwave}
  {Resonators}}.
\bjtitle{Physical Review Applied}
\bvolume{14}(\bissue{4}),
\bfpage{044040}
(\byear{2020}).
\doiurl{10.1103/PhysRevApplied.14.044040}.
Accessed 2023-01-24
\end{barticle}
\endbibitem

\bibitem{Niepce2020}
\begin{barticle}
\bauthor{\bsnm{Niepce}, \binits{D.}},
\bauthor{\bsnm{Burnett}, \binits{J.J.}},
\bauthor{\bsnm{Latorre}, \binits{M.G.}},
\bauthor{\bsnm{Bylander}, \binits{J.}}:
\batitle{Geometric scaling of two-level-system loss in superconducting
  resonators}.
\bjtitle{Superconductor Science and Technology}
\bvolume{33}(\bissue{2}),
\bfpage{025013}
(\byear{2020}).
\doiurl{10.1088/1361-6668/ab6179}.
Accessed 2023-01-13
\end{barticle}
\endbibitem

\bibitem{paper_koblischka1995magneto}
\begin{barticle}
\bauthor{\bsnm{Koblischka}, \binits{M.}},
\bauthor{\bsnm{Wijngaarden}, \binits{R.}}:
\batitle{Magneto-optical investigations of superconductors}.
\bjtitle{Superconductor Science and Technology}
\bvolume{8}(\bissue{4}),
\bfpage{199}
(\byear{1995}).
\doiurl{10.1088/0953-2048/8/4/002}
\end{barticle}
\endbibitem

\bibitem{paper_shaw2018quantitative}
\begin{barticle}
\bauthor{\bsnm{Shaw}, \binits{G.}},
\bauthor{\bsnm{Brisbois}, \binits{J.}},
\bauthor{\bsnm{Pinheiro}, \binits{L.}},
\bauthor{\bsnm{M{\"u}ller}, \binits{J.}},
\bauthor{\bsnm{Blanco~Alvarez}, \binits{S.}},
\bauthor{\bsnm{Devillers}, \binits{T.}},
\bauthor{\bsnm{Dempsey}, \binits{N.}},
\bauthor{\bsnm{Scheerder}, \binits{J.}},
\bauthor{\bparticle{Van~de} \bsnm{Vondel}, \binits{J.}},
\bauthor{\bsnm{Melinte}, \binits{S.}}, \betal:
\batitle{Quantitative magneto-optical investigation of
  superconductor/ferromagnet hybrid structures}.
\bjtitle{Review of Scientific Instruments}
\bvolume{89}(\bissue{2}),
\bfpage{023705}
(\byear{2018}).
\doiurl{10.1063/1.5016293}
\end{barticle}
\endbibitem

\bibitem{paper_blunt1991investigation}
\begin{barticle}
\bauthor{\bsnm{Blunt}, \binits{F.}},
\bauthor{\bsnm{Perry}, \binits{A.}},
\bauthor{\bsnm{Campbell}, \binits{A.}},
\bauthor{\bsnm{Siu}, \binits{R.}}:
\batitle{An investigation of the appearance of positive magnetic moments on
  field cooling some superconductors}.
\bjtitle{Physica C: Superconductivity}
\bvolume{175}(\bissue{5-6}),
\bfpage{539}--\blpage{544}
(\byear{1991}).
\doiurl{10.1016/0921-4534(91)90262-W}
\end{barticle}
\endbibitem

\end{thebibliography}


\end{document}